\documentstyle[12pt]{article}
\begin{document}
%
%%%%%%%%%%%%%%%%%%%%%%%%%%%%%%%%%%%%%%%%%%%%%%%%%%%%%%%%%%%%%%
%
\begin{titlepage}
\begin{raggedleft}
THES-TP 97/05\\
July 1997\\
\end{raggedleft}
\vspace{2em}
\begin{center}
{\Large\bf{ Null Strings in Kerr Spacetime }}
\footnote {Work supported by the PENED No.1768 " Nonlinear theories 
of Gravitation in modern Astrophysics and Microcosmology "}\\
\vspace{3em}
{\Large \bf
P.I. Porfyriadis \footnote {porfyriadis@physics.auth.gr},
D. Papadopoulos \footnote {papadop@helios.astro.auth.gr}}
\vspace{1em}\\
\vspace{2em}
Department of Theoretical Physics, University of
Thessaloniki, Thessaloniki 54006, Greece\\
\end{center}
\vspace{6em}
\begin{abstract}
The null string's equations of motion and constraints in the Kerr
spacetime are given. We assume a generic ansatz for the null strings 
in the Kerr spacetime and we present the resulting solutions in 
quadratures. Some specific string configurations, that follow from 
the generic one, are considered separately. In each case we also 
extract the corresponding solutions in the Schwarzschild spacetime.
\end{abstract}
\end{titlepage}
%
%%%%%%%%%%%%%%%%%%%%%%%%%%%%%%%%%%%%%%%%%%%%%%%%%%%%%%%%%%%%%%%%%
%
\section{Introduction}

The analysis of the classical string equations of motion and
constraints in generic curved backgrounds \cite{VS1} has become an active
area of research \cite{Lar,FSZH,VE,LHG,Lar1,LS1,LS2}. The study of 
classical strings in curved spacetime backgrounds provides us with better 
understanding of the physics of gravitation in the context of string 
theory.

It is well known that the classical evolution
of strings is described by a complicated system of second-order
non-linear coupled partial differential equations \cite{Scherk}. 
In order to properly analyse the motion of classical strings in
gravitation backgrounds we need the exact solutions of these 
equations which can be obtained by choosing a specific ansatz.
 Different forms of symmetric 
ansatz have been proposed. In these, the string evolution equations 
become ordinary differential equations.

In the case of null strings the situation is further simplified since
the null strings are similar to the massless point particles.
Schild \cite{schild} was the first one who introduced 
the notion of a null or
tensionless string, (see also \cite{KL}). The first reference on null 
strings in a curved background can be found in the concluding portion of
Schild's paper. The null strings can be considered as a zero 
approximation with the string tension as the perturbation parameter
\cite{VN,LS}.
Calculations involving null strings in different
metrics have been performed recently \cite{kar,DL,RZ}.

The task of this paper is to discuss the null string evolution in the
Kerr spacetime. In Section 2 we
present the general equations of motion for null
strings in the Kerr spacetime. Next in Section 3 we give
the general solutions in quadratures in the case of the
generic ansatz for the
null strings. In Section 4 we consider the particular case of a 
circular null string and the corresponding solutions in the Schwarzchild 
spacetime. In Section 5 we do the same thing for another interesting 
particular case arising from the generic ansatz.  
Finally in Section 6, we summarize our results and give some
concluding remarks.
%
%%%%%%%%%%%%%%%%%%%%%%%%%%%%%%%%%%%%%%%%%%%%%%%%%%%%%%%%%%%%%%%%%%%%
%
\section{Null Strings in the Kerr Spacetime}

Let us consider a generic background metric given by 
\begin{equation}
ds^2=g_{00}dt^2+g_{11}dr^2+g_{22}d \theta^2+2g_{03}dtd\phi+
g_{33}d \phi^2
\end{equation}
The tensor  $g_{ij}$ for the Kerr background spacetime  is given by
\begin{eqnarray}
g_{00}&=&\frac{2mr}{r^2+a^2 cos^2 \theta}-1  ,\nonumber \\
g_{03}&=&-\frac{2marsin^2 \theta}{r^2+a^2 cos^2 \theta}  ,\nonumber \\
g_{11}&=&\frac{r^2+a^2 cos^2 \theta}{r^2-2mr+a^2}  ,\nonumber \\
g_{22}&=& r^2+a^2 cos^2 \theta  ,\nonumber \\
g_{33}&=&\left(r^2+a^2 +\frac{2mra^2 sin^2 \theta}
{r^2+a^2 cos^2 \theta}\right) sin^2 \theta  ,\nonumber
\end{eqnarray}
where $m, a$ are the mass and the angular momentum per unit mass
respectively. The Kerr black hole has two coordinate singularities
corresponding to the outer and inner horizon
$r_{\pm}=m\pm \sqrt{m^2-a^2}$, for $m^2 > a^2$. We follow the 
notations given in \cite{Dirac,Chan}

Let us consider the null string equations of motion
\begin{equation}
\ddot{X}^{\mu} + \Gamma^{\mu}_{\nu\rho} \dot{X}^{\nu}\dot{X}^{\rho}
= 0   .
\end{equation}
The constraints read as
\begin{eqnarray}
g_{\mu\nu}\dot{X}^{\mu}\dot{X}^{\nu} & = & 0  ,\nonumber \\
&&\\
g_{\mu\nu}\dot{X}^{\mu}X^{\prime\nu} & = & 0 \nonumber  .
\end{eqnarray}
where the overdots and primes denote differentiations with respect
to $\tau$ and $\sigma$ , respectively. We see that for the null
strings we have the null geodesic equations \cite{Dirac, Chan}
supplemented by the
second constraint in the (3). From the second constraint we see
that each point of the string propagates in the direction perpendicular
to the string. We notice that the constraint equations are not
invariant under arbitrary reparametrizations of $\sigma$ and $\tau$.
It is easy to see that equations (2) and constraints (3) are
invariant only under transformations of the form $\tau_1 =f(\tau,\sigma)$
and $\sigma_1 = g(\sigma)$ where $f(\tau,\sigma)$ and $g(\sigma)$ are
arbitrary differentiable functions.

We use the following notation for the coordinates of a null 
string 
\begin{equation}
X^{0} = t(\tau,\sigma), \hspace{0.5cm} X^{1} = r(\tau,\sigma),
\hspace{0.5cm} X^{2} = \theta(\tau,\sigma),
\hspace{0.5cm} X^{3} = \phi(\tau,\sigma)  .
\end{equation}
The equations of motion (2) for a null string moving in the Kerr 
spacetime become now
\begin{eqnarray}
\ddot{t} + 2\Gamma^{0}_{01}\dot{t}\dot{r} +
2\Gamma^{0}_{02}\dot{t}\dot{\theta} +
2\Gamma^{0}_{13}\dot{r}\dot{\phi} +
2\Gamma^{0}_{23}\dot{\theta}\dot{\phi} =0
\nonumber \\
&&\nonumber \\
\ddot{\phi} + 2\Gamma^{3}_{01}\dot{t}\dot{r} +
2\Gamma^{3}_{02}\dot{t}\dot{\theta} +
2\Gamma^{3}_{13}\dot{r}\dot{\phi} +
2\Gamma^{3}_{23}\dot{\theta}\dot{\phi} =0
\nonumber \\
&&\\
\ddot{r} + \Gamma^{1}_{00}\dot{t}^2 +
2\Gamma^{1}_{03}\dot{t}\dot{\phi} +
\Gamma^{1}_{11}\dot{r}^2 +
2\Gamma^{1}_{12}\dot{r}\dot{\theta} +
\Gamma^{1}_{22}\dot{\theta}^2 + \Gamma^{1}_{33}\dot{\phi}^2=0
\nonumber \\
&&\nonumber \\
\ddot{\theta} + \Gamma^{2}_{00}\dot{t}^2 +
2\Gamma^{2}_{03}\dot{t}\dot{\phi} +
\Gamma^{2}_{11}\dot{r}^2 +
2\Gamma^{2}_{12}\dot{r}\dot{\theta} +
\Gamma^{2}_{22}\dot{\theta}^2 + \Gamma^{2}_{33}\dot{\phi}^2=0
\nonumber
\end{eqnarray}
and the constraints (3) can be expressed as follows, using the 
the notation (4):
\begin{eqnarray}
g_{00}\dot{t}^2 + 2g_{03}\dot{t}\dot{\phi} + g_{11}\dot{r}^2 +
g_{22}\dot{\theta}^2 + g_{33}\dot{\phi}^2 &=& 0
\nonumber \\
&& \\
g_{00}\dot{t}t^\prime + g_{03}(\dot{t}\phi^\prime + \dot{\phi}t^\prime)
 + g_{11}\dot{r}r^\prime +
g_{22}\dot{\theta}\theta^\prime + g_{33}\dot{\phi}\phi^\prime &=& 0
\nonumber
\end{eqnarray}
where $\Gamma^{i}_{jk}$ and $ g_{ij} $ are functions from the
$ r \; , \; \theta $.
Making the following substitution: 
\begin{equation}
u=\dot{t} \;\; , \; \; v=\dot{\phi}
\end{equation}
in the (5) we obtain: 
\begin{eqnarray}
\dot{u} + 2(\Gamma^{0}_{01}u + \Gamma^{0}_{13}v)\dot{r}+
2(\Gamma^{0}_{02}u + \Gamma^{0}_{23}v)\dot{\theta} = 0
\nonumber \\
&&\nonumber \\
\dot{v} + 2(\Gamma^{3}_{01}u + \Gamma^{3}_{13}v)\dot{r}+
2(\Gamma^{3}_{02}u + \Gamma^{3}_{23}v)\dot{\theta} = 0
\nonumber \\
&&\\
2g_{11}\ddot{r} +g_{11,1}\dot{r}^2 +
2g_{11,2}\dot{r}\dot{\theta} - g_{22,1}\dot{\theta}^2
-(g_{00,1}u^2+2g_{03,1}uv+g_{33,1}v^2)=0
\nonumber \\
&&\nonumber \\
2g_{22}\ddot{\theta} +g_{22,2}\dot{\theta}^2 +
2g_{22,1}\dot{r}\dot{\theta} - g_{11,2}\dot{r}^2
-(g_{00,2}u^2+2g_{03,2}uv+g_{33,2}v^2)=0
\nonumber
\end{eqnarray}
and the constraints (6) now read 
\begin{eqnarray}
g_{11}\dot{r}^2 + g_{22}\dot{\theta}^2 + g_{00}u^2+2g_{03}uv+
g_{33}v^2= 0
\nonumber \\
&&\\
g_{00}ut^\prime + g_{03}(u\phi^\prime + vt^\prime)
 + g_{11}\dot{r}r^\prime +
g_{22}\dot{\theta}\theta^\prime + g_{33}v\phi^\prime = 0
\nonumber
\end{eqnarray}

The first two equations in (5) cannot be integrated in general,  
although this can be done in the case of Schwarzschild spacetime. 
We will try to investigate the existence of solution when a generic 
ansatz for null strings in the Kerr spacetime is employed.
%
%%%%%%%%%%%%%%%%%%%%%%%%%%%%%%%%%%%%%%%%%%%%%%%%%%%%%%%%%%%%%%%%%
%
\section{Ansatz}

In this section we consider the following generic ansatz for a null 
string in the Kerr spacetime 
\begin{equation}
t =\gamma\sigma+t_{1}(\tau) , \hspace{0.3cm} r = r(\tau), \hspace{0.3cm}
\theta = \theta(\tau), \hspace{0.3cm} \phi = \beta \sigma + \phi_{1}(\tau) .
\end{equation}
which helps us simplify considerably the systems of equations (8) 
and (9).
The variables $\tau$ and $\sigma$ are respectively
 the time-like and space-like coordinates
on the wordsheet  and $\gamma \; , \; \beta $ are constants.
We obtain from (8) the system equations
\begin{equation}
u=C_{1}(\beta g_{33}+\gamma g_{03})\exp\{-G(r,\theta)\}
\end {equation}
\begin{equation}
v=-C_{1}(\gamma g_{00}+\beta g_{03})\exp\{-G(r,\theta)\}
\end {equation}
\begin{equation}
2g_{11}\ddot{r} +g_{11,1}\dot{r}^2 +
2g_{11,2}\dot{r}\dot{\theta} - g_{22,1}\dot{\theta}^2
-(g_{00,1}u^2+2g_{03,1}uv+g_{33,1}v^2)=0
\end {equation}
\begin{equation}
2g_{22}\ddot{\theta} +g_{22,2}\dot{\theta}^2 +
2g_{22,1}\dot{r}\dot{\theta} - g_{11,2}\dot{r}^2
-(g_{00,2}u^2+2g_{03,2}uv+g_{33,2}v^2)=0
\end {equation}
\begin{equation}
g_{11}\dot{r}^2 + g_{22}\dot{\theta}^2 + g_{00}u^2+2g_{03}uv+
g_{33}v^2= 0
\end{equation}
\begin{equation}
(\gamma g_{00}+\beta g_{03})u+(\gamma g_{03}+\beta g_{33})v= 0
\end{equation}
where $C_1$ is constant since $ u $ and $ v $ are function only $\tau$.
The function $G(r,\theta)$ in the exponent have the form 
\begin{equation}
G(r,\theta)=\int g^{00}dg_{00} +2\int g^{03}dg_{03}+
\int g^{33}dg_{33}
\end{equation}

Multiplying (13) by $\dot{r}$, (14) by $\dot{\theta}$, summing the
results together and subtracting the derivative with respect to
$\tau$ of (15) we get
\begin{equation}
\frac{dg}{d\tau}-g\frac{dG}{d\tau}=0
\end{equation}
where $g=g_{00}g_{33}-g_{03}^2$.
After integrating the equation (18) we have 
\begin{equation}
G=ln(g)
\end{equation}
In (17) and (19) we omitted the constants of integration 
which are included in the
definition of $C_1$.

 Now we integrate the equation (14). Multiplying (14) 
by $g_{22}\dot{\theta}$ we have
\begin{equation}
\frac{d}{d\tau}(g_{22}^2 \dot{\theta}^2)+
           K(r,\theta)\frac{d\theta}{d\tau}=0
\end{equation}
where
\begin{equation}
K(r,\theta)=\frac{C_{1}^2}{g^2}[g\frac{\partial g^{\prime} g_{22}}
{\partial \theta} -g^{\prime}g_{22} \frac{\partial g}{\partial \theta}]
\end{equation}
and $g^{\prime}=\gamma^2 g_{00}+2\gamma \beta
g_{03}+\beta^2 g_{33}$.  It is now obvious  
that this equation can be integrated. So finally after having
substituted (19) in (11) and (12), 
we get the following system equations
for the ansatz (10) 
\begin{eqnarray}
\dot{t_{1}}&=&\frac{C_1 (\beta g_{33}+\gamma g_{03})}{g}
\nonumber \\
\dot{\phi_{1}}&=&-\frac{C_1(\gamma g_{00}+ \beta g_{03})}{g}
\\
\dot{r}^2 &=&-\frac{C_{1}^2 g^\prime}{gg_{11}}-
 g^{22}g^{11}[C_2 -a^2 C_{1}^2 \beta^2 sin^2 \theta
 -C_{1}^2 \gamma^2 sin^{-2} \theta]
\nonumber \\
g_{22}^2 \dot{\theta}^2 &=& C_2 -a^2 C_{1}^2 \beta^2 sin^2 \theta
 -C_{1}^2 \gamma^2 sin^{-2} \theta
\nonumber
\end{eqnarray}
 
Notice that if we define the invariant string size as
\begin{equation}
S(\tau)=\int_{0}^{2 \pi} \sqrt{g_{\mu \nu}\dot{X^\mu}\dot{X^\nu}}d\sigma
\end{equation}
we have from equations (22)  and for the Kerr metric
\begin{equation}
S(\tau)=2\pi \sqrt{\gamma^2 g_{00} + 2\gamma\beta g_{03} + \beta^2 g_{33}}
\end{equation}
In the case $\theta=const.$  we have
$C_2 =a^2 C_{1}^{2} \beta^2 sin^{2}\theta+
 C_{1}^2 \gamma^2 sin^{-2} \theta$ 
and from the system of equations (22) follows that 
\begin{eqnarray}
\dot{t_{1}}&=& C_1 (\beta g^{00}-\gamma g^{03})
\nonumber \\
\dot{\phi_{1}}&=&C_1(\beta g^{03}- \gamma g^{33})
\\
\dot{r}^2 &=&- C_{1}^2 g^{11}(\beta^2 g^{00}-2\gamma \beta g^{03}
+\gamma^2 g^{33})
\nonumber
\end{eqnarray}
Now we can integrate the set of equations (25) completely
 and we obtain
\begin{eqnarray}
\pm (t_{1}-t^{0}_{1})=\int_{r}^{r_0} (\beta g^{00}-\gamma g^{03})
[g^{11}(2\gamma \beta g^{03}-\beta^2 g^{00}
-\gamma^2 g^{33})]^{-\frac{1}{2}} dr
\nonumber \\
&&\\
\pm (\phi_1-\phi^{0}_1)=\int_{r}^{r_0} (\beta g^{03}-\gamma g^{33})
[g^{11}(2\gamma \beta g^{03}-\beta^2 g^{00}
-\gamma^2 g^{33})]^{-\frac{1}{2}} dr
\nonumber
\end{eqnarray}
We can also derive from (25)
 the interesting case of a null string moving 
in the Schwarzschild spacetime. The metric can be taken from the 
Kerr's one by putting $a=0$. Then the system of equations for the 
generic ansatz (10) are the following 
\begin{eqnarray}
\dot{t_{1}}&=& C_1 \beta g^{00}
\nonumber \\
\dot{\phi_{1}}&=&-C_1 \gamma g^{33}
\\
\dot{r}^2 &=&C_{1}^{2} \beta^2 -C_{2} g^{22} g^{11}
\nonumber \\
g_{22}^2 \dot{\theta}^2 &=& C_2 - C_{1}^2 \gamma^2 sin^{-2} \theta
\nonumber
\end{eqnarray}
%
% *****************************************************************
%
\section{Circular Null Strings}

It is easy to see that by taking $\gamma=0$ and $\beta=1$ in the 
 generic ansatz (10)  we obtain 
the particular configuration of a circular null string moving in  
Kerr spacetime
\begin{equation}
t = t(\tau) , \hspace{0.3cm} r = r(\tau), \hspace{0.3cm}
\theta = \theta(\tau), \hspace{0.3cm} \phi = \sigma + \phi_{1}(\tau)   .
\end{equation}
The system equation for this particular string configuration can be 
extracted from the more general ones (22), after substituting the values 
of the parameters $\gamma$ and $\beta$. It is checked, however, that 
the same equations can be obtained directly from the equations 
of motion by taking into the account the specific ansatz (28).
\begin{eqnarray}
\dot{t}&=&C_1 g^{00}
\nonumber \\
\dot{\phi_{1}}&=&C_1 g^{03}
\\
\dot{r}^2 &=&-C_{1}^2 g^{00} g^{11}-
 g^{22}g^{11}(C_2 -a^2 C_{1}^2 sin^2 \theta )
\nonumber \\
g_{22}^2 \dot{\theta}^2 &=& C_2 -a^2 C_{1}^2 sin^2 \theta
\nonumber
\end{eqnarray}
Since $\dot{\phi_{1}}$ does not vanish, 
 $\phi$ is a function of both $\sigma$ and $\tau$.
When $\theta=const.$  we have
$C_2 =a^2 C_{1}^2 sin^{2}\theta$ and
\begin{eqnarray}
\pm (t-t_0)=\int_{r}^{r_0} \sqrt{-g^{00}g_{11}} dr
\nonumber \\
&&\\
\pm (\phi_1 -\phi^{0}_1)=
\int_{r}^{r_0} g^{03}\sqrt{-\frac{g_{11}}{g^{00}}} dr
\nonumber
\end{eqnarray}
We return now to the case of  
Schwarzschild spacetime, and consider again the evolution of a 
circular null string. The system equations (29) now takes the form:  
\begin{eqnarray}
\dot{t}&=&C_1 g^{00}
\nonumber \\
\dot{\phi_{1}}&=&0
\\
\dot{r}^2 &=& C_{1}^2 - C_{2} g^{22}g^{11}
\nonumber \\
g_{22}^2 \dot{\theta}^2 &=& C_2
\nonumber
\end{eqnarray}
In this case $\dot{\phi_{1}}$  vanishes and thus 
$\phi$ is a function only of $\sigma$. Consequently, 
in the case of Schwarzchild background metric, 
a possible ansatz for circular null string is given by
\begin{equation}
t=t(\tau)\;,\;\; r=r(\tau)\;,\;\; \theta=\theta(\tau)\;,
\;\; \phi=\sigma
\end{equation}
%
% *****************************************************************
%
\section{Another possible dynamic string configuration}

Let us now interchange the values between the parameters $\gamma$ 
and $\beta$, i.e. let us put $\beta=0$ and $\gamma=1$. We are then led 
 to the following ansatz
\begin{equation}
t = \sigma +t_{1}(\tau) , \hspace{0.3cm} r = r(\tau), \hspace{0.3cm}
\theta = \theta(\tau), \hspace{0.3cm} \phi = \phi(\tau)   .
\end{equation}
Following Section 4, we can substitute in the results of Section 3 
these particular values of the parameters and obtain the following 
system of ordinary differential
equations for this particular string 
configuration \\
\vspace{0.3cm} 
\begin{eqnarray}
\dot{t_{1}}&=&-C_1 \gamma g^{03}
\nonumber \\
\dot{\phi}&=&-C_1 \gamma g^{33}
\\
\dot{r}^2 &=&-C_{1}^2 g^{11} g^{33}-
 g^{22}g^{11}[C_2 - C_{1}^2 \gamma^2 sin^{-2} \theta]
\nonumber \\
g_{22}^2 \dot{\theta}^2 &=& C_2 - C_{1}^2 \gamma^2 sin^{-2} \theta
\nonumber
\end{eqnarray}
When $\theta=const.$ for the Kerr spacetime we have
$C_2 = C_{1}^2 \gamma^2 sin^{-2} \theta$ and
\begin{eqnarray}
\pm (t_{1}-t^{0}_{1})=\int_{r}^{2m} g^{03}
\sqrt{-\frac{g_{11}}{g^{33}}} dr
\nonumber \\
&&\\
\pm (\phi-\phi_{0})=\int_{r}^{2m} \sqrt{-g^{11} g^{33}} dr
\nonumber
\end{eqnarray}
The system equations (35) is meaningful only when $g^{11} g^{33} < 0$.
For the case $\theta =\frac{\pi}{2}$ we have  that
$g^{11} g^{33} = \frac{r-2m}{r^3}$ and from this follows that at the 
equatorial plane we have solution only in the region $r < 2m$.

If the case of Schwarzschild spacetime the equations (34) take 
the form
\begin{eqnarray}
\dot{t_{1}}&=& 0
\nonumber \\
\dot{\phi}&=&-C_1 \gamma g^{33}
\\
\dot{r}^2 &=& -C_{1}^2 g^{11}g^{33}-g^{11}g^{22}
[C_2 - C_{1}^2 \gamma^2 sin^{-2} \theta]
\nonumber \\
g_{22}^2 \dot{\theta}^2 &=& C_2 - C_{1}^2 \gamma^2 sin^{-2} \theta
\nonumber
\end{eqnarray}
It is easy to see that when $\theta=const.$
these equations do not have any solution outside the horizon of the 
 Schwarzschild black hole.
This is the reason why we cannot have an ansatz for the 
string in the form
\begin{equation}
t=\gamma \sigma \;,\;\; r=r(\tau)\;,\;\; \theta=const. \;,
\;\; \phi=\phi(\tau)
\end{equation}
outside the horizon in the case  of the Schwazchild spacetime.
%
%%%%%%%%%%%%%%%%%%%%%%%%%%%%%%%%%%%%%%%%%%%%%%%%%%%%%%%%%%%%%%%%%%%%
%
\section{Conclusion}
\label{Conc}

To conclude, let us now summarize the results we obtained. We have
written down the null string equations of motion and constraints
for the case of the Kerr spacetime. We have been able to construct
the general solutions in quadratures for a generic ansatz for the
null string and we also gave the system of equations 
for the case of Schwarzschild spacetime.

We considered also the particular case of the evolution of a 
circular null string in Kerr spacetime. Here too the solution is 
given in quadratures. The specific case of the Schwarzschild 
spacetime is also discussed and it agrees with other 
authors \cite{kar,DL}. Another null string configuration 
determined by the ansatz (33) is also examined. For the special 
of the equatorial plane it describes the motion of the string 
for $r < 2m $. \\
\vspace{1em}
\\
We wish to thank Argiris Nicolaidis, Jannis Paschalis and Kostas Skenderis 
for useful discussions.
%
%%%%%%%%%%%%%%%%%%%%%%%%%%%%%%%%%%%%%%%%%%%%%%%%%%%%%%%%%%%%%%%%%%%
%

\end{document}